# Call and Put Option Pricing with Discrete Linear Investment Strategy

Niloofar Ghorbani[1], Andrzej Korzeniowski[2]

[1]Department of Mathematical Sciences, High Point University, High Point, USA.
[2]Department of Mathematics, University of Texas at Arlington, Arlington, Texas, USA.
Email: nghorban@highpoint.edu, korzeniowski@uta.edu

## Abstract

We study the Option pricing with linear investment strategy based on discrete time trading of the underlying security, which unlike the existing continuous trading models provides a feasible real market implementation. Closed form formulas for Call and Put Option price are established for fixed interest rates and their extensions to stochastic Vasicek and Hull-White interest rates.

## 1. Introduction

Financial derivatives market over the past few decades led to various generalizations of the classical Black-Scholes model. Notably, a seminal work of Wang et al. ([6] [7]) introduced a dynamic investment strategy in the underlying security for the purpose of hedging the investment risks. It turned out that selling a security proportionally to its dropping price for Put Option and buying the security proportionally to its rising price for Call Option (both under the Black-Scholes model) resulted in lower Option price. Zhang et al. [5] extended the result for Call Option to stochastic interest rates following the Vasicek model. Subsequently Ghorbani and Korzeniowski [2] and [3] obtained the Call Option price with investment strategy for the Cox-Ingersoll-Ross (CIR) interest rates and the Hull-White respectively.

A key assumption in the above modeling relies on the fact that the stock trading by the option holder is done in continuous time. To remedy this drawback and accommodate the actual market implementation, Meng Li et al [1] proposed a discrete time investment strategy for Call Option pricing in the case of non-random constant interest rate and the stock price following the Geometric Fractional Brownian Motion (GFBM).

Novelty of dynamic investment strategies, and their discrete time market implementation presented here, is two-fold. Firstly, unlike in the classical Black-Scholes models where the investor buys options and has no position in the underlying stock throughout the option time horizon, the dynamic investment strategy requires the investor to trade the stock at discrete times associated with the stock price crossing certain levels in the pre-determined price range, whereby lowering the investor's risk, manifested by the lower option price. Secondly, the interest rates are no longer constant and are assumed stochastic.

In this paper the stock price follows the Geometric Brownian Motion (GBM). We first extend the discrete investment strategy for Call Option to random interest rates under the Vasicek and Hull-White models, and subsequently we derive European Put Option price for non-random constant, Vasicek and Hull-White interest rates respectively.

## 2. Call Option

### 2.1. Value Function

It was found in [1] that the Call Option value $V_T$ based on the discrete trading strategy reads

$$V_T = \begin{cases} 0 & S_T < K \\ S_T - K & K \leq S_T < S_1 \\ S_T - \dfrac{\beta S_T K}{N} \sum_{n=1}^{m} \dfrac{1}{K+n\Delta} + \left(\dfrac{\beta m}{N} - 1\right)K & S_m \leq S_T < S_{m+1} \leq S_N \\ S_T - \dfrac{\beta S_T K}{N} \sum_{n=1}^{N} \dfrac{1}{K+n\Delta} + (\beta - 1)K & S_N \leq S_T \end{cases} \quad (2.1)$$

with the investment parameters $\alpha, \beta$, strike price $K$ and the terminal stock price $S_T$. where,

$\alpha$ is the investment strategy index, indicating the stock investment occurs during the period in which the stock price increases from $K$ to $(1+\alpha)K$.

$\beta$ is the maximum value of the stock investment proportion.

The Call Option pricing formula is derived by [1] for the special case of interest rate to be fixed based on the geometric fractional Brownian Motion stock price be-havior and is comparable with classical *B-S* model. In what follows, we compute the Call Option price under the discrete position strategy for the case of non-random, Vasicek and Hull-White interest rate models.

## 2.2. Option Price under Fixed Interest Rates

Suppose $S_T$, the stock price dynamic under the risk-neutral measure be as

$$dS_t = rS_t dt + \sigma_1 S_t dW_{1,t}, \quad S(0) = S_0 > 0, \ 0 \leq t \leq T. \tag{2.2}$$

By Ito formula, the stock price at the maturity time $T$ can be obtained and reads

$$S_T = S_0 e^{\left[\left(r - \frac{1}{2}\sigma_1^2\right)t + \sigma_1 W_{1,T}\right]}. \tag{2.3}$$

Based on the value function $V_T$ in (2.1), the Call Option price under risk-neutral measure can be evaluated as

$$C = e^{-rT} E[V_T] \tag{2.4}$$

where,

$$\begin{aligned} E[V_T] &= \int_K^{S_1} (S_T - K) f(S_T) dS_T \\ &+ \int_{S_m}^{S_{m+1}} \left[\left(1 - \frac{\beta K}{N} \sum_{n=1}^{m} \frac{1}{K + n\Delta}\right) S_T + \left(\frac{\beta m}{N} - 1\right) K\right] f(S_T) dS_T \\ &+ \int_{S_N}^{+\infty} \left[\left(1 - \frac{\beta K}{N} \sum_{n=1}^{N} \frac{1}{K + n\Delta}\right) S_T + (\beta - 1) K\right] f(S_T) dS_T \end{aligned} \tag{2.5}$$

with

$$\begin{aligned} I_1 &= \int_K^{S_1} (S_T - K) f(S_T) dS_T \\ I_2 &= \int_{S_m}^{S_{m+1}} \left[\left(1 - \frac{\beta K}{N} \sum_{n=1}^{m} \frac{1}{K + n\Delta}\right) S_T + \left(\frac{\beta m}{N} - 1\right) K\right] f(S_T) dS_T \\ I_3 &= \int_{S_N}^{+\infty} \left[\left(1 - \frac{\beta K}{N} \sum_{n=1}^{N} \frac{1}{K + n\Delta}\right) S_T + (\beta - 1) K\right] f(S_T) dS_T. \end{aligned} \tag{2.6}$$

Calculating the three integrals $I_1$, $I_2$ and $I_3$ with a similar method to the derivations in [2] and [3] using the probability density function of $\ln S_T$, with mean and variance as follow

$$\begin{aligned} \mu = E[\ln S_T] &= E\left[\ln S_0 + \left(r - \frac{1}{2}\sigma_1^2\right) T + \sigma_1 W_{1,T}\right] \\ &= \ln S_0 + \left(r - \frac{1}{2}\sigma_1^2\right) T \end{aligned} \tag{2.7}$$

$$\sigma_1^2 = Var\left[\ln S_0 + \left(r - \frac{1}{2}\sigma_1^2\right)T + \sigma_1 W_{1,T}\right] \tag{2.8}$$
$$= \sigma_1^2 T$$

gives the Call Option price.

$$\begin{aligned}
I_1 &= \int_K^{S_1} (S_T - K) f(S_T) dS_T \\
&= \int_{\ln K}^{\ln S_1} (e^y - K) f(e^y) e^y dy \\
&= \int_{\ln K}^{\ln S_1} (e^y - K) \frac{1}{\sqrt{2\pi}\sigma} e^{-\frac{1}{2}\frac{(y-\mu)^2}{\sigma^2}} dy \\
&= \int_{\ln K}^{\ln S_1} \frac{1}{\sqrt{2\pi}\sigma} e^y e^{-\frac{1}{2}\frac{(y-\mu)^2}{\sigma^2}} dy - K \int_{\ln K}^{\ln S_1} \frac{1}{\sqrt{2\pi}\sigma} e^{-\frac{1}{2}\frac{(y-\mu)^2}{\sigma^2}} dy \\
&= \int_{\frac{\ln K - \mu}{\sigma}}^{\frac{\ln S_1 - \mu}{\sigma}} \frac{1}{\sqrt{2\pi}} e^{\mu+\sigma z} e^{-\frac{1}{2}z^2} dz - K \int_{\frac{\ln K - \mu}{\sigma}}^{\frac{\ln S_1 - \mu}{\sigma}} \frac{1}{\sqrt{2\pi}} e^{-\frac{1}{2}z^2} dz \\
&= \frac{1}{\sqrt{2\pi}} \int_{\frac{\ln K - \mu}{\sigma}}^{\frac{\ln S_1 - \mu}{\sigma}} e^{\mu+\frac{1}{2}\sigma_2} e^{-\frac{1}{2}(z-\sigma)^2} dz - K \frac{1}{\sqrt{2\pi}} \int_{\frac{\ln K - \mu}{\sigma}}^{\frac{\ln S_1 - \mu}{\sigma}} e^{-\frac{1}{2}z^2} dz \\
&= e^{\mu+\frac{1}{2}\sigma_2} \left[ N\left(\frac{\ln S_1 - \mu - \sigma^2}{\sigma}\right) - N\left(\frac{\ln K - \mu - \sigma^2}{\sigma}\right) \right] \\
&\quad - K \left[ N\left(\frac{\ln S_1 - \mu}{\sigma}\right) - N\left(\frac{\ln K - \mu}{\sigma}\right) \right].
\end{aligned} \tag{2.9}$$

and

$$\begin{aligned}
I_2 &= \int_{S_m}^{S_{m+1}} \left[ \left(1 - \frac{\beta K}{N} \sum_{n=1}^{m} \frac{1}{K + n\Delta}\right) S_T + \left(\frac{\beta m}{N} - 1\right) K \right] f(S_T) dS_T \\
&= \int_{S_m}^{S_{m+1}} \left(1 - \frac{\beta K}{N} \sum_{n=1}^{m} \frac{1}{K + n\Delta}\right) S_T f(S_T) dS_T + \int_{S_m}^{S_{m+1}} \left(\frac{\beta m}{N} - 1\right) K f(S_T) dS_T \\
&= \left(1 - \frac{\beta K}{N} \sum_{n=1}^{m} \frac{1}{K + n\Delta}\right) \int_{S_m}^{S_{m+1}} e^y f(e^y) e^y dy + \left(\frac{\beta m}{N} - 1\right) K \int_{S_m}^{S_{m+1}} f(e^y) e^y dy \\
&= \frac{1}{\sqrt{2\pi}\sigma} \left(1 - \frac{\beta K}{N} \sum_{n=1}^{m} \frac{1}{K + n\Delta}\right) \int_{S_m}^{S_{m+1}} e^y e^{-\frac{1}{2}\frac{(y-\mu)^2}{\sigma^2}} dy \\
&\quad + \frac{1}{\sqrt{2\pi}\sigma} \left(\frac{\beta m}{N} - 1\right) K \int_{S_m}^{S_{m+1}} e^{-\frac{1}{2}\frac{(y-\mu)^2}{\sigma^2}} dy \\
&= \frac{1}{2\pi} \left(1 - \frac{\beta K}{N} \sum_{n=1}^{m} \frac{1}{K + n\Delta}\right) \int_{\frac{S_m - \mu}{\sigma}}^{\frac{S_{m+1} - \mu}{\sigma}} e^{\mu+z\sigma} e^{-\frac{1}{2}z^2} dz + \frac{1}{2\pi} \left(\frac{\beta m}{N} - 1\right) K \int_{\frac{S_m - \mu}{\sigma}}^{\frac{S_{m+1} - \mu}{\sigma}} e^{-\frac{1}{2}z^2} dz \\
&= e^{\mu+\frac{1}{2}\sigma^2} \left(1 - \frac{\beta K}{N} \sum_{n=1}^{m} \frac{1}{K + n\Delta}\right) \left[ N\left(\frac{\ln S_{m+1} - \mu - \sigma^2}{\sigma}\right) - N\left(\frac{\ln S_m - \mu - \sigma^2}{\sigma}\right) \right] \\
&\quad + \left(\frac{\beta m}{N} - 1\right) K \left[ N\left(\frac{\ln S_{m+1} - \mu}{\sigma}\right) - N\left(\frac{\ln S_m - \mu}{\sigma}\right) \right].
\end{aligned} \tag{2.10}$$

$$\begin{aligned}
I_3 &= \int_{S_N}^{+\infty}\left[\left(1-\frac{\beta K}{N}\sum_{n=1}^{N}\frac{1}{K+n\Delta}\right)S_T+(\beta-1)K\right]f(S_T)dS_T \\
&= \left(1-\frac{\beta K}{N}\sum_{n=1}^{N}\frac{1}{K+n\Delta}\right)\int_{\ln S_N}^{+\infty}e^y f(e^y)e^y\,dy+(\beta-1)K\int_{\ln S_N}^{+\infty}f(e^y)e^y\,dy \\
&= \left(1-\frac{\beta K}{N}\sum_{n=1}^{N}\frac{1}{K+n\Delta}\right)\frac{1}{\sqrt{2\pi}\sigma}\int_{\ln S_N}^{+\infty}e^y e^{-\frac{1}{2}\frac{(y-\mu)^2}{\sigma^2}}dy \\
&\quad +(\beta-1)K\frac{1}{\sqrt{2\pi}\sigma}\int_{\ln S_N}^{+\infty}e^{-\frac{1}{2}\frac{(y-\mu)^2}{\sigma^2}}dy \\
&= e^{\mu+\frac{1}{2}\sigma^2}\left(1-\frac{\beta K}{N}\sum_{n=1}^{N}\frac{1}{K+n\Delta}\right)\left[N\left(\frac{\mu+\sigma^2-\ln S_N}{\sigma}\right)\right] \\
&\quad +(\beta-1)KN\left(\frac{\mu-\ln S_N}{\sigma}\right).
\end{aligned} \quad (2.11)$$

Thus, from formulas (2.4) to (2.11), the Option Price under fixed interest rate can be obtained by

$$\begin{aligned}
C &= e^{-rT}E[V_T] \\
&= e^{-rT}e^{\mu+\frac{1}{2}\sigma^2}[N(d_1)-N(d_2)] \\
&\quad -Ke^{-rT}[N(d_3)-N(d_4)] \\
&\quad +e^{-rT}e^{\mu+\frac{1}{2}\sigma^2}\left(1-\frac{\beta K}{N}\sum_{n=1}^{m}\frac{1}{K+n\Delta}\right)[N(d_5)-N(d_6)] \\
&\quad +e^{-rT}\left(\frac{\beta m}{N}-1\right)[N(d_7)-N(d_8)] \\
&\quad +e^{-rT}e^{\mu+\frac{1}{2}\sigma^2}\left(1-\frac{\beta K}{N}\sum_{n=1}^{m}\frac{1}{K+n\Delta}\right)N(d_9) \\
&\quad +e^{-rT}K(\beta-1)N(d_{10})
\end{aligned} \quad (2.12)$$

where,

$$d_1 = \frac{\ln S_1 - \mu - \sigma^2}{\sigma} \qquad d_6 = \frac{\ln S_m - \mu - \sigma^2}{\sigma}$$

$$d_2 = \frac{\ln K - \mu - \sigma^2}{\sigma} \qquad d_7 = \frac{\ln S_{m+1} - \mu}{\sigma}$$

$$d_3 = \frac{\ln S_1 - \mu - \sigma^2}{\sigma} \qquad d_8 = \frac{\ln S_m - \mu}{\sigma}$$

$$d_4 = \frac{\ln K - \mu}{\sigma} \qquad d_9 = \frac{\mu + \sigma^2 - \ln S_N}{\sigma}$$

$$d_5 = \frac{\ln S_{m+1} - \mu - \sigma^2}{\sigma} \qquad d_{10} = \frac{\mu - \ln S_N}{\sigma}$$

and

$$\mu = \ln S_0 + \left(\mu - \frac{1}{2}\sigma_1^2\right)T$$

$$\sigma^2 = \sigma_1^2 T.$$

**Remark.** The Option price under fixed interest rate with the discrete trading strategy was derived in the geometric fractional Brownian motion setting for stock price evolution in [1]. For the sake of further analysis we derived it in this paper directly rather than as a special case for stock price geometric Brownian motion.

### 2.3. Option Price under Stochastic Interest Rates

### 2.3.1. Vasicek Model

Solution to the Vasicek interest rate model [4] can be used to obtain the Call Option price $C$ according to the following formula

$$C = P(0,T) E^T [V_T] \tag{2.13}$$

where

$$S_T = S_0 e^{\int_0^T \left(r_s - \frac{1}{2}\sigma_1^2\right)ds + \int_0^T \sigma_1 dW_{1,t}} \tag{2.14}$$

and $P(0,T)$ is the price of the zero-coupon bond. Also, the mean $\mu_T$ and variance $\sigma_T^2$ in the case of Vasicek has been calculated under the *T*-forward measure by Zhang [5]. Thus, the expectation of $V_T$ under the *T*-forward measure based on the discrete dynamic strategy can be obtained similar to (2.12) and is as follows

$$C = P(0,T)e^{\mu_T+\frac{1}{2}\sigma_T^2}\left[N(d_1)-N(d_2)\right]$$
$$-KP(0,T)\left[N(d_3)-N(d_4)\right]$$
$$+P(0,T)e^{\mu_T+\frac{1}{2}\sigma_T^2}\left(1-\frac{\beta K}{N}\sum_{n=1}^{m}\frac{1}{K+n\Delta}\right)\left[N(d_5)-N(d_6)\right] \quad (2.15)$$
$$+P(0,T)\left(\frac{\beta m}{N}-1\right)\left[N(d_7)-N(d_8)\right]$$
$$+P(0,T)e^{\mu_T+\frac{1}{2}\sigma_T^2}\left(1-\frac{\beta K}{N}\sum_{n=1}^{m}\frac{1}{K+n\Delta}\right)N(d_9)$$
$$+P(0,T)K(\beta-1)N(d_{10})$$

where,

$$d_1 = \frac{\ln S_1 - \mu_T - \sigma_T^2}{\sigma_T} \qquad d_6 = \frac{\ln S_m - \mu_T - \sigma_T^2}{\sigma_T}$$
$$d_2 = \frac{\ln K - \mu_T - \sigma_T^2}{\sigma_T} \qquad d_7 = \frac{\ln S_{m+1} - \mu_T}{\sigma_T}$$
$$d_3 = \frac{\ln S_1 - \mu_T - \sigma_T^2}{\sigma_T} \qquad d_8 = \frac{\ln S_m - \mu_T}{\sigma_T}$$
$$d_4 = \frac{\ln K - \mu_T}{\sigma_T} \qquad d_9 = \frac{\mu_T + \sigma^2 - \ln S_N}{\sigma_T}$$
$$d_5 = \frac{\ln S_{m+1} - \mu_T - \sigma_T^2}{\sigma_T} \qquad d_{10} = \frac{\mu_T - \ln S_N}{\sigma_T}$$

and

$$\mu_T = \ln S_0 - \frac{\sigma_1^2}{2}T - r_0\frac{1-e^{-aT}}{a} + \left(\frac{\theta}{a}-\frac{\sigma_2^2}{a^2}\right)\left[T-\frac{1-e^{-aT}}{a}\right]+\frac{\sigma_2^2}{2a}\frac{1-e^{-aT}}{a}$$
$$\sigma_T^2 = \frac{\sigma_2^2}{a^2}\left[T-2\frac{1-e^{-aT}}{a}+\frac{1-e^{-2aT}}{2a}\right]+\sigma_1^2 T$$

with the zero-coupon bond price as

$$P(0,T) = \exp\left\{\left[\frac{1-e^{-aT}}{a}-T\right]\left(\frac{\theta}{a}-\frac{\sigma_2^2}{2a^2}\right)-\frac{\sigma_2^2\frac{1-e^{-aT}}{a}}{4a}\right\}e^{-r_0\frac{1-e^{-aT}}{a}}. \quad (2.16)$$

### 2.3.2. Hull-White Model

We derived the Option price under Hull-White interest rates model [3] using the solution to the Hull-White *SDE* obtained in [3]. Since our derivation is under *T*-forward measure, we need to consider the zero-coupon bond price *P(0,T)* and the mean $\mu_T$ and Variance $\sigma_T^2$ for this model under the *T*-forward measure. The Option price in this case reads (2.15), where

$$\mu_T = \ln S_0 - \frac{\sigma_1^2}{2}T + r_0 \frac{1-e^{-aT}}{a} - \frac{\sigma_2^2}{a}\left[\frac{e^{-2aT}\left((2aT-3)e^{2aT}+4e^{aT}-1\right)}{2a^2}\right]$$

$$+ \int_0^T e^{-at} \int_0^t \theta(s)e^{as}\,dsdt$$

$$\sigma_T^2 = \frac{\sigma_2^2}{a^2}\left[T - 2\frac{1-e^{-aT}}{a} + \frac{1-e^{-2aT}}{2a}\right] + \sigma_1^2 T$$

and the zero-coupon bond price is

$$P(0,T) = e^{\frac{r_0(e^{-aT}-1)}{a} - \int_0^T e^{-as}\int_0^s \theta(u)e^{au}\,duds + \frac{\sigma_2^2}{2a^2}\left[T+\frac{1-e^{-2aT}}{2a} - \frac{2}{a}(1-e^{-aT})\right]}. \tag{2.17}$$

## 3. Put Options

### 3.1. Value Function

The Put Option contains a specified capital amount $A$ and holding of $Q$ initially. When the stock price drops from $K$ to $(K-\delta)$, $\delta \geq 0$ the Option holder sells $\beta_0 A$ proportion of stocks. Parameter $\beta_0$ is called the initial capital utilization coefficient which is a constant between $0$ and $1$. The Option holder linearly adjusts the capital utilization to decrease the holding if the price continues to fall until it reaches $(1-\alpha)(K-\delta)$ and the total capital spending reaches $\beta A$ with $\beta < \ldots < \beta_1 < \beta_0$. The Option holder will only sell when the stock price hits a series of equally spaced points $S_n$, where $\{S_n : S_n = K - n\Delta,\ n=1,2,\ldots,N\}$, with $\Delta$ a positive constant, the price distance for two consecutive trading actions and $N$ the total number of trades within the capital valid period. The strategy parameters $\alpha$ (investment index) and $\beta$ (minimum capital utilization), both positive numbers, illustrate the maximum amount of capital tradable is $\beta_0 A$ when the price reaches $(1-\alpha)(K-\delta)$. The strategy assumes the Option holder will evenly $\left(\frac{\beta A}{N}\right)$ distribute the capital over each of the potential trades corresponding to $S_n$, $n=1,2,\ldots,N$ by selling $\frac{\beta A}{NS_n}$ shares of stock.

In the classical European Options, the Option writer will sustain a loss of the amount $L$ as the stock price falls below $K$

$$L = Q(K-S) = \frac{A}{K}(K-S). \tag{3.1}$$

For the purpose of adaptive hedging Option, the investor is required to sell a proportion of underlying stock in order to hedge the risk. In a discrete position strategy, the Option holder's income $R$ based on such transactions throughout the Option valid period can be calculated as,

$$R = \frac{\beta A}{NS_n}(S_n - S) \tag{3.2}$$

when the stock price falls from $S_n$ to $S$, with $S_N \leq S_{m+1} < S < S_n$.

The Option holder makes a cumulative income $R(S)$ as

$$R(S) = \sum_{n=1}^{m} \frac{\beta A}{NS_n}(S_n - S) \tag{3.3}$$

Thus, the total loss taken by the Option writer can be obtained as

$$\begin{aligned}L(S) &= \frac{A}{K}(K - S) - \sum_{n=1}^{m} \frac{\beta A}{NS_n}(S_n - S) \\ &= \frac{A}{K}(K - S) - \frac{\beta A}{N}\sum_{n=1}^{m}\frac{(S_n - S)}{S_n} \\ &= \frac{A}{K}(K - S) - \frac{\beta A}{N}\sum_{n=1}^{m}\frac{S_n}{S_n} + \frac{\beta AS}{N}\sum_{n=1}^{m}\frac{1}{S_n} \\ &= A - \frac{AS}{K} - \frac{\beta Am}{N} + \frac{\beta AS}{N}\sum_{n=1}^{m}\frac{1}{K - n\Delta} \\ &= \left(1 - \frac{\beta m}{N}\right)A - \frac{AS}{K} + \frac{\beta AS}{N}\sum_{n=1}^{m}\frac{1}{K - n\Delta}\end{aligned} \tag{3.4}$$

however, for $S \leq S_N$, the Option writer's loss $L$ is

$$L(S) = (1 - \beta)A - \frac{AS}{K} + \frac{\beta AS}{N}\sum_{n=1}^{m}\frac{1}{K - n\Delta}. \tag{3.5}$$

Therefore, the following function represents the Option writer's Loss based on the stock price $S$

$$L(S) = \begin{cases} (1-\beta)A - \frac{AS}{K} + \frac{\beta AS}{N}\sum_{n=1}^{m}\frac{1}{K - n\Delta} & S \leq S_N \\ \left(1 - \frac{\beta m}{N}\right)A - \frac{AS}{K} + \frac{\beta AS}{N}\sum_{n=1}^{m}\frac{1}{K - n\Delta} & S_N \leq S_{m+1} < S \leq S_m \\ \frac{A}{K}(K - S) & S_1 < S \leq K \\ 0 & S > K \end{cases} \tag{3.6}$$

Consequently, the intrinsic value function $V(S)$ is derived by dividing the loss function by the number of shares of stock within the Option depending on the stock finishing price $S_T$

$$V(S_T) = \begin{cases} (1-\beta)K - S_T + \dfrac{\beta S_T K}{N}\sum_{n=1}^{m}\dfrac{1}{K-n\Delta} & S_T \leq S_N \\ \left(1-\dfrac{\beta m}{N}\right)K - S_T + \dfrac{\beta S_T K}{N}\sum_{n=1}^{m}\dfrac{1}{K-n\Delta} & S_N \leq S_{m+1} < S_T \leq S_m. \\ K - S_T & S_1 < S_T \leq K \\ 0 & S_T > K \end{cases} \quad (3.7)$$

## 3.2. Option Price under Fixed Interest Rates

Considering the assumptions of section (3.1) and the intrinsic value function $V(S_T)$ (3.7), the Put Option price under risk-neutral measure is evaluated as

$$P = e^{-rT} E[V(S_T)]$$

where,

$$\begin{aligned} E[V(S_T)] &= \int_0^{S_N}\left[(1-\beta)K - S_T + \dfrac{\beta S_T K}{N}\sum_{n=1}^{m}\dfrac{1}{K-n\Delta}\right]f(S_T)dS_T \\ &+ \int_{S_{m+1}}^{S_m}\left[\left(1-\dfrac{\beta m}{N}\right)K - S_T + \dfrac{\beta S_T K}{N}\sum_{n=1}^{m}\dfrac{1}{K-n\Delta}\right]f(S_T)dS_T \quad (3.8) \\ &+ \int_{S_1}^{K}(K-S_T)f(S_T)dS_T \end{aligned}$$

calculating the three integrals $I_1$, $I_2$ and $I_3$ similar to the previous sections and using the probability density function of $\ln S_T$, whose mean and variance has been obtained in (2.7) and (2.8) gives the Put Option price.

$$\begin{aligned} I_1 &= \int_0^{S_N}\left[(1-\beta)K - S_T + \dfrac{\beta S_T K}{N}\sum_{n=1}^{m}\dfrac{1}{K-n\Delta}\right]f(S_T)dS_T \\ &= \int_0^{S_N}(1-\beta)K f(S_T)dS_T - \int_0^{S_N}\left(1-\dfrac{\beta K}{N}\sum_{n=1}^{m}\dfrac{1}{K-n\Delta}\right)S_T f(S_T)dS_T \\ &= (1-\beta)K\int_{-\infty}^{\ln S_N}f(e^y)e^y dy - \left(1-\dfrac{\beta K}{N}\sum_{n=1}^{m}\dfrac{1}{K-n\Delta}\right)\int_{-\infty}^{\ln S_N}e^y f(e^y)e^y dy \\ &= (1-\beta)K\dfrac{1}{\sqrt{2\pi}\sigma}\int_{-\infty}^{\ln S_N}e^{-\frac{1}{2}\frac{(y-\mu)^2}{\sigma^2}}dy - \left(1-\dfrac{\beta K}{N}\sum_{n=1}^{m}\dfrac{1}{K-n\Delta}\right)\dfrac{1}{\sqrt{2\pi}\sigma}\int_{-\infty}^{\ln S_N}e^y e^{-\frac{1}{2}\frac{(y-\mu)^2}{\sigma^2}}dy \\ &= (1-\beta)K\dfrac{1}{\sqrt{2\pi}}\int_{-\infty}^{\frac{\ln S_N - \mu}{\sigma}}e^{-\frac{1}{2}z^2}dz - \left(1-\dfrac{\beta K}{N}\sum_{n=1}^{m}\dfrac{1}{K-n\Delta}\right)\dfrac{1}{\sqrt{2\pi}}\int_{-\infty}^{\frac{\ln S_N - \mu}{\sigma}}e^{\mu+z\sigma}e^{-\frac{1}{2}z^2}dz \\ &= (1-\beta)K\dfrac{1}{\sqrt{2\pi}}\int_{-\infty}^{\frac{\ln S_N - \mu}{\sigma}}e^{-\frac{1}{2}z^2}dz - \left(1-\dfrac{\beta K}{N}\sum_{n=1}^{m}\dfrac{1}{K-n\Delta}\right)\dfrac{1}{\sqrt{2\pi}}\int_{-\infty}^{\frac{\ln S_N - \mu}{\sigma}}e^{\mu+z\sigma}e^{-\frac{1}{2}(z-\sigma)^2}dz \end{aligned}$$

$$= (1-\beta) K\, N\left(\frac{\ln S_N - \mu}{\sigma}\right) - \left(1 - \frac{\beta K}{N}\sum_{n=1}^{m}\frac{1}{K - n\Delta}\right) N\left(\frac{\ln S_N - \mu - \sigma^2}{\sigma}\right) \quad (3.9)$$

and

$$\begin{aligned}
I_2 &= \int_{S_{m+1}}^{S_m}\left[\left(1 - \frac{\beta m}{N}\right)K - S_T + \frac{\beta S_T K}{N}\sum_{n=1}^{m}\frac{1}{K - n\Delta}\right] f(S_T)\, dS_T \\
&= \left(1 - \frac{\beta m}{N}\right) K \int_{S_{m+1}}^{S_m} f(e^y) e^y\, dy - \left(1 - \frac{\beta K}{N}\sum_{n=1}^{m}\frac{1}{K - n\Delta}\right)\int_{S_{m+1}}^{S_m} e^y f(e^y) e^y\, dy \\
&= \left(1 - \frac{\beta m}{N}\right) K \frac{1}{\sqrt{2\pi}\sigma}\int_{S_{m+1}}^{S_m} e^{-\frac{1}{2}\frac{(y-\mu)^2}{\sigma^2}}\, dy - \left(1 - \frac{\beta K}{N}\sum_{n=1}^{m}\frac{1}{K - n\Delta}\right) \frac{1}{\sqrt{2\pi}\sigma}\int_{S_{m+1}}^{S_m} e^y e^{-\frac{1}{2}\frac{(y-\mu)^2}{\sigma^2}}\, dy \\
&= \left(1 - \frac{\beta m}{N}\right) K \frac{1}{\sqrt{2\pi}}\int_{\frac{S_{m+1}-\mu}{\sigma}}^{\frac{S_m - \mu}{\sigma}} e^{-\frac{1}{2}z^2}\, dz - \left(1 - \frac{\beta K}{N}\sum_{n=1}^{m}\frac{1}{K - n\Delta}\right)\frac{1}{\sqrt{2\pi}}\int_{\frac{S_{m+1}-\mu}{\sigma}}^{\frac{S_m - \mu}{\sigma}} e^{\mu + \frac{1}{2}\sigma^2} e^{-\frac{1}{2}(z - \sigma)^2}\, dz \\
&= \left(1 - \frac{\beta m}{N}\right) K \left[ N\left(\frac{\ln S_m - \mu}{\sigma}\right) - N\left(\frac{\ln S_{m+1} - \mu}{\sigma}\right)\right] \\
&\quad - \left(1 - \frac{\beta K}{N}\sum_{n=1}^{m}\frac{1}{K - n\Delta}\right)\left[ N\left(\frac{\ln S_m - \mu - \sigma^2}{\sigma}\right) - N\left(\frac{\ln S_{m+1} - \mu - \sigma^2}{\sigma}\right)\right]
\end{aligned}$$

(3.10)

similarly,

$$\begin{aligned}
I_3 &= \int_{S_1}^{K}(K - S_T) f(S_T)\, dS_T \\
&= K\int_{\ln S_1}^{\ln K} f(e^y) e^y\, dy - \int_{\ln S_1}^{\ln K} e^y f(e^y) e^y\, dy \\
&= K \frac{1}{\sqrt{2\pi}\sigma}\int_{\ln S_1}^{\ln K} e^{-\frac{(y-\mu)^2}{\sigma^2}}\, dy - \frac{1}{\sqrt{2\pi}\sigma}\int_{\ln S_1}^{\ln K} e^y e^{-\frac{(y-\mu)^2}{\sigma^2}}\, dy \\
&= K\left[N\left(\frac{\ln K - \mu}{\sigma}\right) - N\left(\frac{\ln S_1 - \mu}{\sigma}\right)\right] - e^{\mu + \frac{1}{2}\sigma^2}\left[N\left(\frac{\ln K - \mu - \sigma^2}{\sigma}\right) - N\left(\frac{\ln S_1 - \mu - \sigma^2}{\sigma}\right)\right]
\end{aligned}$$

(3.11)

### 3.3. Option Price under Stochastic Interest Rates

### 3.3.1. Vasicek Model

The price of Put Option $P$ with the underlying stock price dynamic and interest rate following the Vasicek model reads

$$P = P(0,T) E^T[V_T]$$

where $P(0,T)$ is the zero-coupon bond price (2.16) with $\mu_T$, $\sigma_T^2$ the mean and variance under $T$-forward measure. Thus,

$$\begin{aligned}
P &= P(0,T) E^T[V_T] \\
&= P(0,T) K (1-\beta) N(d_1) \\
&\quad - P(0,T)\left(1 - \frac{\beta K}{N} \sum_{n=1}^{m} \frac{1}{K - n\Delta}\right) N(d_2) \\
&\quad + P(0,T)\left(1 - \frac{\beta m}{N}\right) K \left[N(d_3) - N(d_4)\right] \\
&\quad - P(0,T)\left(1 - \frac{\beta K}{N} \sum_{n=1}^{m} \frac{1}{K - n\Delta}\right) \left[N(d_5) - N(d_6)\right] \\
&\quad + P(0,T)\left[N(d_7) - N(d_8)\right] \\
&\quad - P(0,T) e^{\mu + \frac{1}{2}\sigma^2} \left[N(d_9) - N(d_{10})\right]
\end{aligned} \qquad (3.12)$$

where,

$$d_1 = \frac{\ln S_N - \mu_T}{\sigma_T} \qquad d_6 = \frac{\ln S_{m+1} - \mu_T - \sigma_T^2}{\sigma_T}$$

$$d_2 = \frac{\ln S_N - \mu_T - \sigma_T^2}{\sigma_T} \qquad d_7 = \frac{\ln K - \mu_T}{\sigma_T}$$

$$d_3 = \frac{\ln S_m - \mu_T}{\sigma_T} \qquad d_8 = \frac{\ln S_1 - \mu_T}{\sigma_T}$$

$$d_4 = \frac{\ln S_{m+1} - \mu_T}{\sigma_T} \qquad d_9 = \frac{\ln K - \mu_T - \sigma_T^2}{\sigma_T}$$

$$d_5 = \frac{\ln S_m - \mu_T - \sigma_T^2}{\sigma_T} \qquad d_{10} = \frac{\ln S_1 - \mu_T - \sigma_T^2}{\sigma_T}$$

and

$$\mu_T = \ln S_0 - \frac{\sigma_1^2}{2} T - r_0 \frac{1 - e^{-aT}}{a} + \left(\frac{\theta}{a} - \frac{\sigma_2^2}{a^2}\right)\left[T - \frac{1 - e^{-aT}}{a}\right] + \frac{\sigma_2^2}{2a} \frac{1 - e^{-aT}}{a}$$

$$\sigma_T^2 = \frac{\sigma_2^2}{a^2}\left[T - 2\frac{1 - e^{-aT}}{a} + \frac{1 - e^{-2aT}}{2a}\right] + \sigma_1^2 T.$$

### 3.3.2. Hull-White Model

The Put Option price under the extended Vasicek, Hull-White model reads (3.12) where the mean $\mu_T$ and variance $\sigma_T^2$ under the $T$-forward measure are

$$\mu_T = \ln S_0 - \frac{\sigma_1^2}{2}T + r_0 \frac{1-e^{-aT}}{a} - \frac{\sigma_2^2}{a}\left[\frac{e^{-2aT}\left((2aT-3)e^{2aT}+4e^{aT}-1\right)}{2a^2}\right]$$

$$+\int_0^T e^{-at}\int_0^t \theta(s)e^{as}\,ds\,dt$$

$$\sigma_T^2 = \frac{\sigma_2^2}{a^2}\left[T - 2\frac{1-e^{-aT}}{a} + \frac{1-e^{-2aT}}{2a}\right] + \sigma_1^2 T$$

where the zero-coupon bond price is

$$P(0,T) = e^{\frac{r_0(e^{-aT}-1)}{a} - \int_0^T e^{-as}\int_0^s \theta(e^{au})\,du\,ds + \frac{\sigma_2^2}{2a^2}\left[T + \frac{1-e^{-2aT}}{2a} - \frac{2}{a}(1-e^{-aT})\right]}. \quad (3.13)$$